\documentclass{article}
\usepackage[english]{babel}
\usepackage[margin=1in]{geometry}
\usepackage{multicol}
\usepackage[math]{blindtext}
\usepackage[linesnumbered, ruled, vlined]{algorithm2e}
\usepackage{mathtools}
\usepackage{caption}
\usepackage{subcaption}

\usepackage{graphicx, balance}
\usepackage[figuresright]{rotating}

\title{\textbf{Optimal Two-Tier Forecasting Power Generation Model in Smart Grids}}

\begin{document}

{\centering

{\bfseries\Large Optimal Two-Tier Forecasting Power Generation Model\\ in Smart Grids\bigskip}

Kianoosh G. Boroojeni\textsuperscript{1} ,Shekoufeh Mokhtari\textsuperscript{1} , M.H. Amini\textsuperscript{2} , and S.S. Iyengar \textsuperscript{1}\\
  {\itshape
\textsuperscript{1}School of Computing and Information Sciences, Florida International University, Miami, Florida, USA \\
\textsuperscript{2}Department of Electrical and Computer Engineering, Carnegie Mellon University, Pittsburgh, PA, USA \\

  }
}
\begin{abstract}
There has been an increasing trend in the electric power system from a centralized generation-driven grid to a more reliable, environmental friendly, and customer-driven grid. One of the most important issues which the designers of smart grids need to deal with is to forecast the fluctuations of power demand and generation in order to make the power system facilities more flexible to the variable nature of renewable power resources and demand-side. This paper proposes a novel two-tier scheme for forecasting the power demand and generation in a general residential electrical gird which uses the distributed renewable resources as the primary energy resource. The proposed forecasting scheme has two tiers: long-term demand/generation forecaster which is based on Maximum-Likelihood Estimator (MLE) and real-time demand/generation forecaster which is based on Auto-Regressive Integrated Moving-Average (ARIMA) model. The paper also shows that how bulk generation improves the adequacy of proposed residential system by canceling-out the forecasters estimation errors which are in the form of Gaussian White noises.  \\\\
{\bf Keywords :} Adequacy Analysis, ARIMA Model, Forecasting Model, Maximum Likelihood Estimation, Smart Grids
\end{abstract}

\section{INTRODUCTION}
In recent years, increasing awareness about environmental issues and sustainable energy supply introduced modern power system, called smart grid (SG), to upgrade conventional power system by utilizing novel technologies. There are many influential elements in the SG which helps power grid to achieve a more reliable, sustainable, efficient and secure level, such as distributed renewable resources (DRRs), advanced metering infrastructure (AMI), energy storage devices, electric vehicles, demand response programs, energy efficiency programs, and home area networks (HANs) \cite{h1,h2}. Furthermore, recent advances in deploying communication networks in SG provide two-way communication between utility and electricity consumers and improve market efficiency \cite{h3}. In a related context, conventional generation resources mostly use fossil fuel as their energy source which is a major environmental concern. To overcome this problem, SG will experience a high penetration of DRRs which has two main advantages: 1) cost-effective because the main energy source is free (wind energy, sunlight, etc), 2) produce no hazardous pollution. Additionally, DRR utilization helps power system to become dispersed. Therefore, not only SG is more distributed than conventional power system but power generation units are trying to implement green-based energy resources \cite{h4}. In order to reduce the variability of wind power generation, this reference proposes coordination of wind power plant and flexible load\cite{j1}.

One of the most challenging issues in future power system design and implementation is the flexibility of power system devices to adapt the stochastic nature of demand and generation \cite{h5,h6}. In other words, high penetration of DRRs, such as wind power and photovoltaics, is not sufficient to achieve an acceptable level of reliability in terms of adequate supply of electricity demand; for instance the output power of wind generators requires excessive cost to manage intermittency \cite{h7}. Ref \cite{j2} proposes a Consensus+Innovations distributed control approach for the distributed energy resources. Consequently, there is a foremost obligation to develop an accurate forecasting method to predict the power generation of intermittent DRRs.

Based on US Energy Information Administration (EIA) assessment, energy demand will increase by 56\% from 2010 to 2040. This astonishing consumption growth is driven by economic development \cite{h8}. Additionally, on the demand side, customers demand depends on many factors and there are many studies performed in order to achieve an accurate forecast methodology. Load forecast uncertainty plays a pivotal role on power system studies such as loss estimation, reliability evaluation, and generation expansion planning. Demand forecasting methods including, but not limited to, fuzzy logic approach, artificial neural network, linear regression, transfer functions, Bayesian statistics, judgmental forecasting, and grey dynamic models \cite{h9}.

Considering all of the above-mentioned issues, including uncertainty of demand and generation, forecasting errors and DRR intermittent generation profile, proposing an accurate demand/generation forecasting scheme is definitely required to achieve a more reliable and secure power grid. In other words, the purpose of this paper is to propose a framework in which the SG customers are satisfied in terms of supplying their demand reliably, independent from the wide variation of DRRs' generation amount.

\subsection{Related Works}
Utilizing green power generation units, DRRs, requires electricity demand/generation forecasting which are addressed in recent studies. In \cite{h5}, deferrable demand is used to compensate the uncontrollable and hard-to-predict fluctuations of DRRs. As a result, green-based power generation units can utilize the flexibility of the customers to meet demand appropriately. They introduced an efficient solution using stochastic dynamic programming implement their method.  
Moreover, consumer participation in generation side is modeled in term of demand response.  Implementation of demand response programs brings many advantages for the SG: 1) customer participation in generating power, 2) transmission lines congestion management, and 3) reliability improvement. In \cite{h11}, a flexible demand response model is proposed. This model is useful for evaluation customer's reactions to electricity price and incentives. They defined a strategy success index to evaluate the feasibility of each scenario \cite{h11}.

Additionally, Hernandez \textit{et al.} performed a comprehensive survey on power grid demand forecasting methods considering SG elements. This study classifies load forecasting based on forecasting horizons: very short term load forecasting (from seconds to minutes), short term load forecasting (from hours to weeks), medium and long term load forecasting (from months to years). The authors also classified forecasting methods according to the objective of forecast: one value forecasting (next minute's load, next year's load), and multiples values forecasting (such as peak load, average load, load profile) \cite{h9}.

Smart load management studies also require load/generation forecasting to efficiently balance load and generation in a near-real time manner. In \cite{h12}, a multi-agent based load management framework is introduced. This approach considered renewable resources and responsive demand to achieve an acceptable level of load-generation balance. For a broader treatment on this, there are some other researches on this topic that considered electric energy dispatch in presence of DRRs \cite{h13,h14}. For a broader treatment on this, please see \cite{r2,r5,r9,windforecasting,reliableperspective,r10,r11,r12,r13,book1,book2,confpaper1}.

\subsection{Our Contribution}
In this paper, we propose a novel hybrid (two-tier) scheme for forecasting the power demand and generation in a residential electrical gird. The grid has expanded over a city consisting of a number of communities, Distributed Renewable Resources (DRR), and some bulk generations back-up plan. Our forecasting scheme has two tiers: long-term demand/generation forecaster which is based on Maximum-Likelihood Estimator (MLE) and real-time forecaster which is based on Auto-Regressive Integrated Moving-Average (ARIMA) model (see \cite{mle1,mle2,mle3,mle4} for related work). In the long-term forecaster, we use the classification of historical demand/generation data to build our estimator; while in the real-time one, we predict the time series of power demand/generation using a discrete feedback control system which gets feedback from short-term previous values. We show that how the bulk generators can improve the adequacy of our residential system by canceling-out the forecasters estimation errors which are in the form of Gaussian White noises.

The rest of this paper is organized as follows. Section 2 represents a general framework of the problem. In Section 3 we discuss our proposed two-tier forecasting scheme in both long-term and real-time time horizons. A far-reaching adequacy analysis framework is presented in Section 4. Finally, summary and outlook are given in Section 5.
\section{PROBLEM SPECIFICATION}
\begin{figure}[t!]
\centering
\begin{subfigure}[b]{.58\textwidth}
\centering
\includegraphics[width=\textwidth]{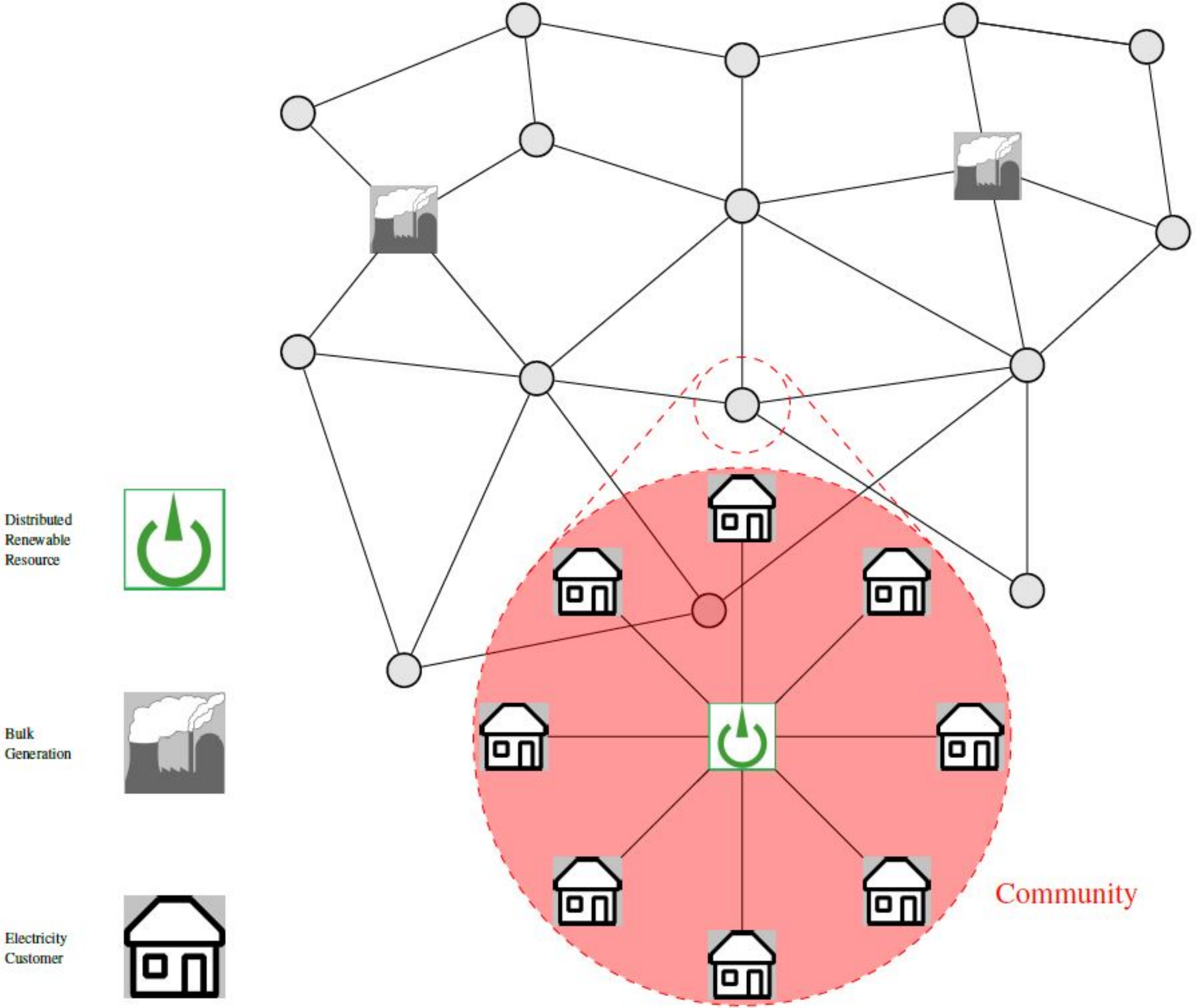}
\caption{the whole network.}
\label{fig:powerdisnet}
\end{subfigure}
\\
\begin{subfigure}[b]{.58\textwidth}
\centering
\includegraphics[width=\textwidth]{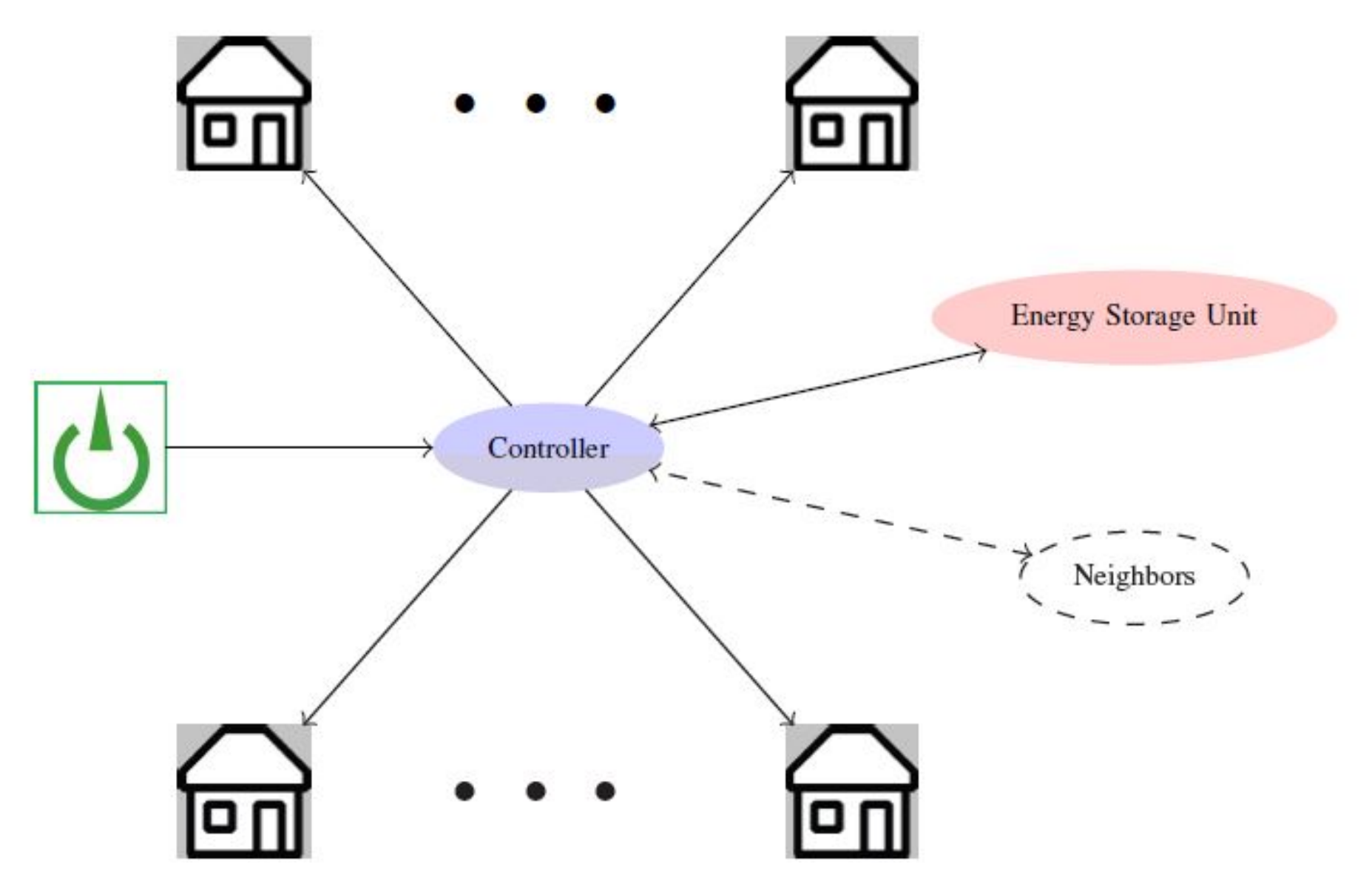}
\caption{More details inside a community. Arrows represent energy flow.}
\label{fig:storage}
\end{subfigure}
\caption{Schematic representation of the electric power distribution network.}
\label{fig:double}
\end{figure}

Consider a network of communities in a city. In each community, there are a number of customers and a \textit{distributed renewable resource}\footnote{An industrial facility for the generation of electric power. The power generator of a distributed renewable resource use renewable energy sources.} which supplies the energy needed by the customers in the community. Moreover, there are a few power plants which are outside the communities and scattered over the city to help the distributed renewable resources generate electricity \textit{on demand}. Existence of these extra plants improves the performance of our electrical distribution system. We refer to these extra power plants as \textit{bulk generators}. The electric energy generated by these generators can be transferred to each community in the city through the network of communities schematically represented in Figure \ref{fig:powerdisnet}. This figure shows a network representation of our described model. As you see, each community has some connected neighbors so that it can trade electricity with them if it is necessary. We assume that any two neighbors are connected via a two-way transmission power line.

\begin{figure}[t!]
\centering
\includegraphics[width=.68\textwidth]{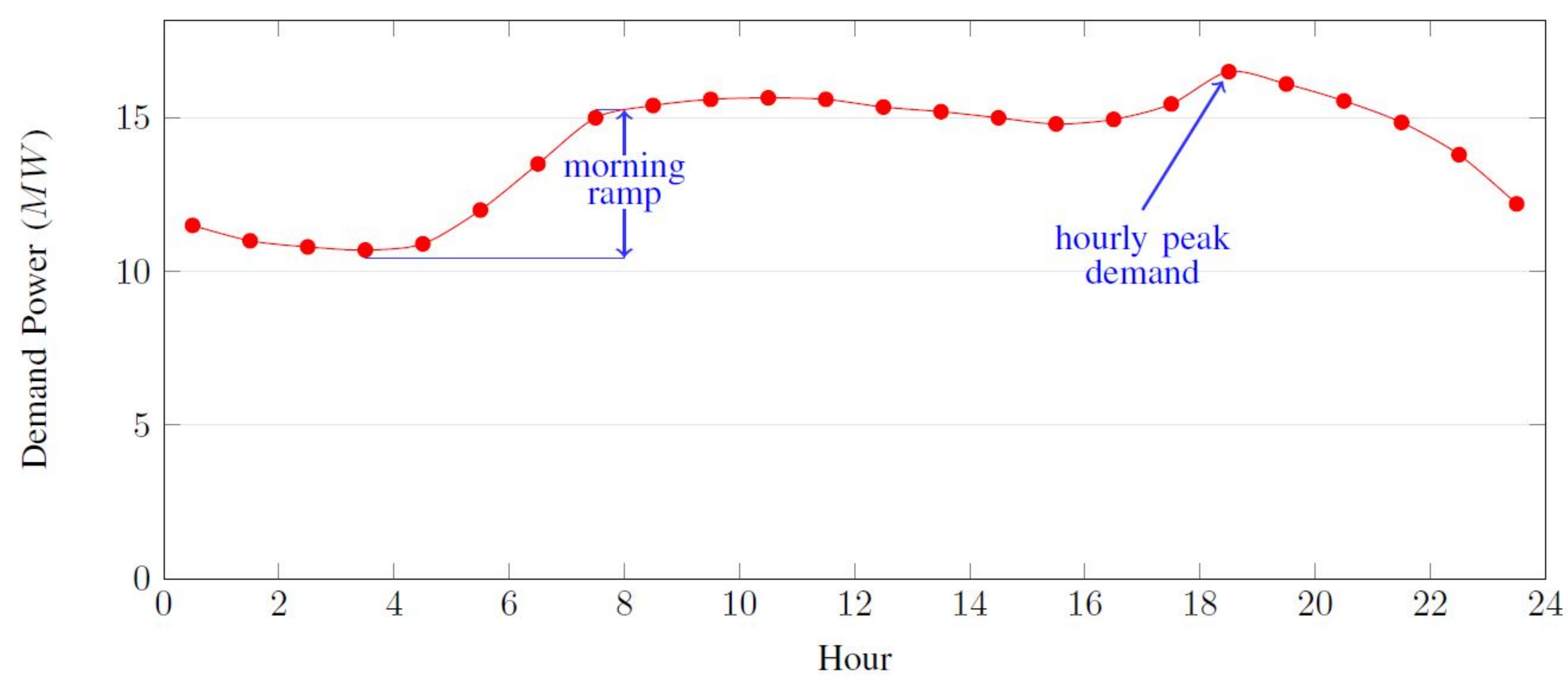}
\caption{Daily demand profile of New-England in 2012.}
\label{fig:neweng}
\end{figure}

Consider some community in our example. Each customer located in the community has some amount of electricity demand which varies from time to time. Let $\mathcal{D}(q,t)$ denote the total electricity demand of all of the customers in the $ q^{th} $ community at given moment $t$. Note that $\mathcal{D}(q,t)$ is the instantaneous power that have to be supplied at time $t$. To illustrate, see Figure \ref{fig:neweng} which specifies the daily expected value of $\mathcal{D}(q,t)$ in New-England in 2012. Additionally, we use $\mathcal{G}(q,t)$ to represent the total amount of electric power generated by the DRR located in the community at given time $t$.

Figure \ref{fig:storage} shows more details of a community. As you see, there is a controller unit called Local Load Management Unit (LLMU) which controls the energy flow inside the community. Furthermore, the controller may participate in the energy distribution of other communities by forwarding the flow received from its neighbors. Assuming that the instantaneous generated power ($\mathcal{G}(q,t)$) of the DRR exceeds $\mathcal{D}(q,t)$ in a period of time, the energy flow of size $\mathcal{G}(q,t)$ from the DRR is divided into two parts: a power flow of size $\mathcal{D}(q,t)$ moves towards the customers and the other $\mathcal{G}(q,t)-\mathcal{D}(q,t)$ units is stored by the energy storage unit embedded in each community. On the other hand, if the value of $\mathcal{G}(q,t)$ becomes lower than $\mathcal{D}(q,t)$ in some time interval, there will be two main energy flows in the community to satisfy the customers demand: one is originated from the DRR; the other of magnitude $ \mathcal{D}(q,t)-\mathcal{G}(q,t) $ is from the storage unit. Moreover, in the case that $\mathcal{D}(q,t)>\mathcal{G}(q,t)$ and there is no enough amount of energy in the storage unit to compensate the shortage of DRR generation, there has to be another flow from the neighbors towards the community of customers. 

\section{OUR PROPOSED HYBRID FORECASTING SCHEME}
In this section, we focus on how to forecast the power demand and generation in short/long-term. At the first subsection, we propose a Maximum-Likelihood Estimator for long-term foresting which is crucial in the process of low-cost energy flow management and also is a basis for real-time forecasting of power demand/generation. In the following subsection, two estimation models will be proposed for real-time forecasting of demand and generation. In the first one which is a two-tier hybrid model (based on MLE and AR), we assume that the forecast random processes are stationary in few hours; however, the second model (which is an ARIMA) is more appropriate for the case that the forecast random processes doesn't show stationary behaviors even in few hours.

\subsection{Long-Term Forecasting}
We assume that there are a set of customers $ C $ distributed over an area. By partitioning the area into $ n $ disjoint parts $ A_1,A_2,\ldots,A_n $, we obtain a corresponding partition of set $ C $: $ C_1,C_2,\ldots,C_n $ (communities of customers). The demand values of every subset $ C_q $ has been measured every $ u $ units in time period $ [0,Tu] $ for some integer $ T $ and real value $ u $. 

Assume that a year is divided into $ m $ parts (school time, Christmas holidays, Summer break, etc) based on the similarity of electricity usage pattern. We partition time interval $ [0,Tu] $ into $ m $ subsets: $ I_{1}, I_{2}, \ldots, I_{m}$ such that set $ I_i $ contains the $ i^{th} $ part of every year belonging to $ [0,Tu] $. Additionally, every set $ I_{i} $ is divided to two parts: weekends $ I_{i1} $ and business days $ I_{i2} $. Moreover, assuming that a day is divided into $ d $ parts (again based on the similarity of electricity usage pattern during the day), we partition every interval $ I_{ij} $ into $ I_{ij1}, I_{ij2},\ldots, I_{ijd} $.

In addition, assume that we have the historical weather data in every area $ A_q $ over period $ [0,Tu] $. Considering that $W $ denotes the set of different weather conditions, we partition time interval $ I_{ijk} $ in the following form for every area $ A_q $:
\begin{equation}
I_{ijk}=\bigcup\limits_{w\in W}I_{ijk}^{(w,q)}
\label{eq:first}
\end{equation}
for every  $i=1,2,\ldots m$, $j=1,2$, $k=1,2,\ldots,d$, $q=1,2,\ldots,n$. Note that in Equation \ref{eq:first}, $ I_{ijk}^{(w,q)} $ specifies the subset of $I_{ijk}$ such that the weather condition in area $ A_q $ and time $t\in I_{ijk}^{(w,q)} $ is $ w $.

Now, assuming that interval $ [0,Tu] $ contains $ \delta $ days, let $ D_v $ specifies the $ v^{th} $ day of time interval $ [0,Tu] $ for every $ v=1,2,\ldots,\delta $. Additionally, consider $ \mathcal{D}(q,\tau) $ as the power demanded by the set of customers $ C_q $ measured at moment $ u\tau $  (for every $ \tau=0,1,\ldots, T $). For every interval $ I_{ijk}^{(w,q,v)}=I_{ijk}^{(w,q)}\cap D_v $, if $ I_{ijk}^{(w,q,v)}\ne\emptyset $, we specify five parameters: $X_{1}^{(q)}$ which is the number of years passed since $ t=0 $ (till interval $ I_{ijk}^{(w,q,v)}$), $ X_{2}^{(q)}$ which is the number of weeks passed since the begining of the $ i^{th} $ paritition of a year, $ X_{3}^{(q)}$ which is the number of days passed since the begining of the $ j^{th} $ partition of a week, $X_{4}^{(q)}$ is the temperature in area $ A_q $ and time interval $ I_{ijk}^{(w,q,v)}$, and
\begin{equation}
y^{(q)}=\dfrac{\sum\limits_{u\tau\in I_{ijk}^{(w,q,v)}}\mathcal{D}(q,\tau)}{\sum\limits_{u\tau\in I_{ijk}^{(w,q,v)}}1}~,
\end{equation}
where $ y^{(q)} $ specifies the average power demanded by the set of customers $ C_q $ in time interval $ I_{ijk}^{(w,q,v)} $. 

For every subset $ I_{ijk}^{(w,q)}\subset [0,Tu] $, we construct a maximum-likelihood estimator for the dependent variable $ y^{(q)} $ based on the following linear model:
\begin{equation}
\begin{split}
\widehat{y}^{(q)} &=[1~X_{1}^{(q)}~X_{2}^{(q)}~X_{3}^{(q)}~X_{4}^{(q)}][\widehat{\beta_0}~\widehat{\beta_1}~\ldots~\widehat{\beta_4}]^T\\
&+N(0,\widehat{\sigma^2})~.
\end{split}
\label{eq:estimatorMLE}
\end{equation}

Considering that condition $ I_{ijk}^{(w,q,v)}\ne \emptyset$ is only true for $ v=v_1,v_2,\ldots,v_p $, we obtain that $
Y^{(q)}=X^{(q)}\beta+\varepsilon~~~~~~~~~~\forall q=1,2,\ldots,n
$
such that $Y^{(q)}=[y_i^{(q)}]_{p\times 1}$, $\beta=[\beta_{i-1}]_{5\times 1}$, $\varepsilon=[\varepsilon_i]_{p\times 1}$, and $X^{(q)}=[1~X_{1}^{(q)}~X_{2}^{(q)}~X_{3}^{(q)}~X_{4}^{(q)}]$. Symbol $ y_{\iota}^{(q)} $ specifies the average power demanded by the set of customers $ C_q $ in time interval $ I_{ijk}^{(w,q,v_{\iota})} $; moreover, $X_{\iota 1},\ldots,X_{\iota 4}$ denote the parameters on which $  y_{\iota}^{(q)} $ is dependent (for every $ \iota=1,2,\ldots,p $).

Using the maximum-likelihood method for the linear model mentioned in Equation \ref{eq:estimatorMLE}, we obtain that:
\begin{equation}
\widehat{\beta}_{ML}=({X^{(q)}}^T{X^{(q)}})^{-1}{X^{(q)}}^T{Y^{(q)}}^T~,
\end{equation}
\begin{equation}
\widehat{\varepsilon}_{ML}=\Big({Y^{(q)}}^T-{X^{(q)}}\hat{\beta}_{ML}\Big)\sim {N}(0,\widehat{\sigma^2}_{ML}I)~,
\end{equation}
and
\begin{equation}
\begin{split}
\widehat{\sigma^2}_{ML}&=\Big(Y^{(q)}-X^{(q)}\widehat{\beta}_{ML}\Big)^T\\
&\times \Big(Y^{(q)}-X^{(q)}\widehat{\beta}_{ML}\Big)/p~.
\end{split}
\end{equation}

Note that the ML estimator specified in Equation \ref{eq:estimatorMLE} can forecast the average power demand in an interval of few hours. However, by using the estimator repetitively and for different intervals $ I_{ijk}^{(w,q)} $, we can forecast the average power demand for longer time; however, the variance of error will increase respectively\footnote{addition of $ b $ i.i.d. jointly normally distributed random variables of variance $ \sigma^2 $ is also a normal variable of variance $ b\sigma^2 $.}.

Additionally, the similar estimation model can be made for power generation. The only difference is that we don't need to partition a week into two parts. Moreover, we have to partition a year into small parts based on the similarity of power generation pattern.
\subsection{Real-Time Forecasting}
In the previous subsection, we partitioned the interval $ [0,Tu] $ into $ \Theta(md|W|) $ subsets in the form of $ I_{ijk}^{(w,q)} $ (for every set of customers $ C_q $). Additionally, for every subset $ I_{ijk}^{(w,q)} $, a maximum likelihood estimator was constructed to estimate the average power demanded by customers $ C_q $ in time interval $ I_{ijk}^{(w,q)}\cap D_v $. 

Our ultimate goal in this section is to construct an estimator for the value of power demanded by set of customers $ C_q $ in moment $ t=\tau u $ (for some integer value $ \tau $) based on ARIMA$ (a,0,0) $ model with drift $ -\mu^{(q)} $:
\begin{equation}
\Big(1-\sum\limits_{l=1}^{a}\phi_lL^l\Big)(\mathcal{D}(q,\tau)-\mu^{(q)})=\varepsilon_{\tau}~,
\label{eq:arimaestimation}
\end{equation}
where $ \mu^{(q)} $ is the average of demand value $ \mathcal{D}(q,t) $ in time interval $ t\in I_{ijk}^{(w,q,v)} $ which is estimated by Equation \ref{eq:estimatorMLE}, $ \varepsilon_{\tau} $ is a white noise of variance $ \sigma^2 $, $ \tau u\in I_{ijk}^{(w,q,v)}$ for some $ i,j,k,w, v $, and $ L $ is the \textit{lag operator}: $
L(\mathcal{D}(q,\tau))=\mathcal{D}(q,\tau-1)$. By replacing $\mu^{(q)}$ with $ \widehat{y}^{(q)}+\varepsilon' $ where $ \varepsilon'\sim N(0,\widehat{\sigma^2}_{ML}) $, we obtain that:
\begin{equation}
\begin{split}
\mathcal{D}(q,\tau)&=\underbrace{(\sum\limits_{l=1}^{a}\phi_l -1)\widehat{y}^{(q)}+\sum\limits_{l=1}^{a}\mathcal{D}(q,\tau-l)}\\
&~~~~~~~~~~~~~~~~\mbox{estimated value}\\
&+\underbrace{\varepsilon_{\tau}+(\sum\limits_{l=1}^{a}\phi_l -1)\varepsilon'}~.\\
&~~~~~\mbox{estimation error}
\end{split}
\end{equation}
Note that Equation \ref{eq:arimaestimation} works only if random process $ \mathcal{D}(q,t) $ shows stationary behavior; otherwise, we need to use the model with moving average. In fact, assuming that process $ \mathcal{D}(q,t) $ is not stationary, ARIMA$ (a,1,0) $ is much better for short-term forecasting:
\begin{equation}
\Big(1-\sum\limits_{l=1}^{a}\phi_lL^l\Big)(1-L)\mathcal{D}(q,\tau)=\varepsilon_{\tau}~.
\end{equation}
Consequently, we obtain that:
\begin{equation}
\begin{split}
\mathcal{D}(q,\tau)&=(\phi_1+1)\mathcal{D}(q,\tau-1)\\
&+\sum\limits_{l=2}^{a}(\phi_l-\phi_{l-1})\mathcal{D}(q,\tau-l)\\
&-\phi_a\mathcal{D}(q,\tau-a-1)+\varepsilon_{\tau}~.
\end{split}
\label{eq:demarimaestimation}
\end{equation}
As you see, ARIMA$ (a,1,0) $ model forecasts the demand value using its $ (a+1) $ previous values with a white noise error. 

In addition, the power generation of the $ g^{th} $ generator can also be forecast using ARIMA$ (a',1,0) $. Assuming that $ \mathcal{G}(g,t) $ specifies the instantaneous power generated by the $ g^{th} $ generator at moment $ t $, we have:
\begin{equation}
\Big(1-\sum\limits_{l=1}^{a'}\phi'_lL^l\Big)(1-L)\mathcal{G}(g,\tau)=\varepsilon'_{\tau}~;
\end{equation}
or equivalently,
\begin{equation}
\begin{split}
\mathcal{G}(g,\tau)&=(\phi'_1+1)\mathcal{G}(g,\tau-1)\\
&+\sum\limits_{l=2}^{a'}(\phi'_l-\phi'_{l-1})\mathcal{G}(g,\tau-l)\\
&-\phi'_{a'}\mathcal{G}(g,\tau-a'-1)+\varepsilon'_{\tau}~.
\end{split}
\label{eq:genarimaestimation}
\end{equation}

In the following section, we analyze the adequacy of the electricity system based on ARIMA$ (a,1,0) $ forecasting model.
\section{ADEQUACY ANALYSIS}
By assumption, we consider the maximum security for our electrical facilities (like wires). Henceforth,
the system reliability in our discussion refers to the system adequacy. In order to analyze the
system adequacy, we need to use the forecasting models of instantaneous demand and generation presented in the previous section:$
\mathcal{D}(q,\tau)=\widehat{\mathcal{D}}(q,\tau)+D_{\tau}$ and $\mathcal{G}(q,\tau)=\widehat{\mathcal{G}}(q,\tau)+G_{\tau}$ such that $ \widehat{\mathcal{D}}(q,\tau) $ and $ \widehat{\mathcal{G}}(q,\tau) $ are obtained by the ARIMA estimators specified in Equations \ref{eq:demarimaestimation} and \ref{eq:genarimaestimation}; additionally, $D_t$ and $G_t$ are two independent \textit{Gaussian white noises} of the following covariance functions (regarding the Central-Limit theorem, the estimation errors of the instantaneous demand and generation are Gaussian processes):
$\mbox{cov}(D_s,D_t)=\sigma_d^2\cdot\delta(s-t)$ and
$\mbox{cov}(G_s,G_t)=\sigma_g^2\cdot\delta(s-t)$.

Now, assume that community $ C_q $ uses the $ q^{th} $ renewable power plant (DRR) to satisfy its demand. Assuming that at given time $t$, community $ C_q $ has stored $\mathcal{S}(q,t)$ units of energy, we obtain that
$
\mathcal{S}(q,t)=\int_{0}^{t}(\mathcal{G}(q,t')-\mathcal{D}(q,t'))\mbox{d}t'+s_q
$ for every $ t\ge 0 $ such that $s_q$ is the initial stored energy in the community. By replacing the generation and demand functions with their equivalent random processes, we obtain that $\mathcal{S}(q,t)=\widehat{\mathcal{S}}(q,t)-\mathcal{W}_t$ where $
\widehat{\mathcal{S}}(q,t)=\int_{0}^{t}(\widehat{\mathcal{G}}(q,t')-\widehat{\mathcal{D}}(q,t'))\mbox{d}t'+s_q
$ and
$
\mathcal{W}_t=\int_0^t(D_{t'}-G_{t'})\mbox{d}t'
$. Since $G_t$ and $D_t$ are two independent Gaussian white noises, $\mathcal{W}_t$ is a Wiener process of variance $(\sigma_g^2+\sigma_d^2)$ and covariance function $ cov(\mathcal{W}_s,\mathcal{W}_t)=\min\{s,t\}\cdot(\sigma_g^2+\sigma_d^2)$.
Moreover, it is easy to show that the expected value of the stored energy at given time $t$ is $\widehat{\mathcal{S}}(q,t)$.

According to the above analysis, the amount of stored energy $\mathcal{S}(q,t)$ is equal to the summation of deterministic amount $\widehat{\mathcal{S}}(q,t)$ and the scaled Wiener process $(-\mathcal{W}_t)$. In the rest of our analysis, we assume that the expected value of the stored energy never becomes less than the initial amount of energy ($s_q$); \textit{i.e.} $\widehat{\mathcal{S}}(q,t)\ge s_q$ for every $t\ge 0$. This condition can be held by providing sufficient DRRs for every community (which is designed based on long-term forecasting of power demand and generation).

\begin{figure}[t!]
\centering
\includegraphics[width=.66\textwidth]{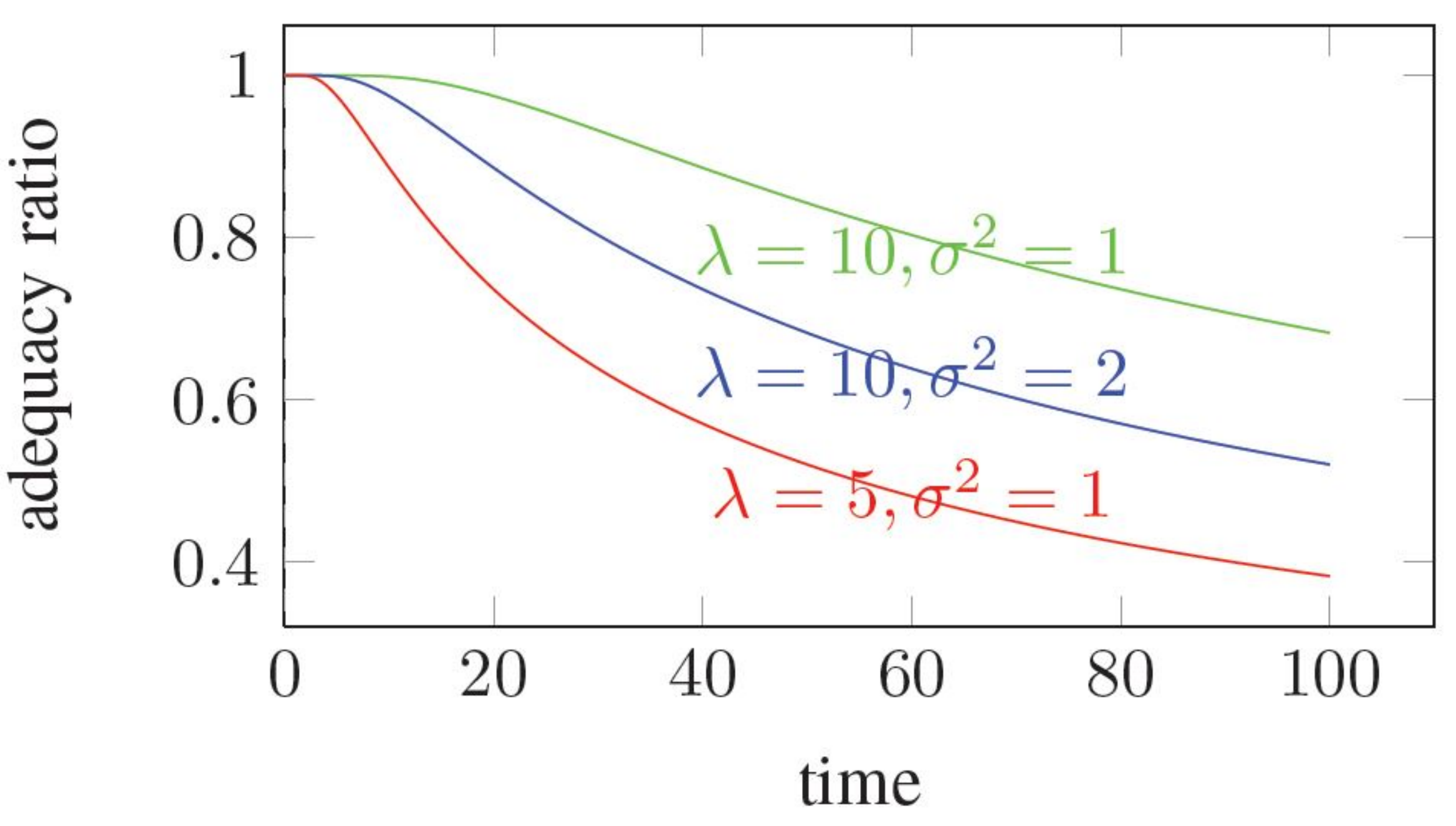}
\caption{Lower bound of the adequacy ratio of a community for three different values of $\lambda$ and $\sigma^2$.}
\label{fig:erffunction}
\end{figure}

Here, we define the system \textit{adequacy ratio} ($ \rho_q(0,t) $) for the $ q^{th} $ community as the probability that the actual stored energy $\mathcal{S}(q,t')$ doesn't meet the low-threshold $(s_q-\lambda)$ for some $\lambda\in [0,s_q]$ and every $t'\in [0,t]$. So, $ \rho_q(0,t)=\textbf{\mbox{Pr}}\big[\forall t'\le t:\mathcal{S}(q,t')>s_q-\lambda\big]\ge\textbf{\mbox{Pr}}\big[M_t<\lambda\big]$ where $M_t$ is the running maximum process corresponding to the scaled Wiener process $\mathcal{W}_t$. So, regarding the characteristics of the running maximum process, we conclude that $\rho_q(0,t)\ge \mbox{erf}\big(\dfrac{\lambda}{\sqrt{2t\sigma^2}}\big)$ such that $\sigma^2=\sigma_g^2+\sigma_d^2$. In other words, we assume that if $\mathcal{S}(q,t)\le s_q-\lambda$ (the stored energy becomes lower than some threshold in the $ q^{th} $ community), the consumers demand will not be satisfied anymore. Figure \ref{fig:erffunction} shows how the lower-bound of the reliability ratio  changes as parameters $\lambda$ and $\sigma^2$ get different values. 

\begin{algorithm}
\SetAlgoLined
\SetKw{KwAnd}{and}
\SetKw{KwTrue}{true}
\SetKw{KwFalse}{false}
\SetKw{KwCon}{continue}
\KwIn{Community Index $ q $ \& time $ t $}

	\If{$\widehat{\mathcal{S}}(q,t+1)\le s_q-\lambda$\hspace{10mm}}{
		Ask the bulk generations for $s_q-\widehat{\mathcal{S}}(q,t+1)$ units of energy\;
	}
		\If{$ \mathcal{D}(q,t)> \mathcal{G}(q,t)$}{
			Create an energy flow of size $ \mathcal{D}(q,t)-\mathcal{G}(q,t) $ originated from the storage unit toward the customers.\;
		}\Else{
			Divide the energy flow originated from the DRR into two branches: one toward the customers and the other toward the storage unit.
		}

\caption{\textsc{LocalLoadManagementUnit}}
\label{alg:decmakinprocess}
\end{algorithm}
As you see in Figure \ref{fig:erffunction}, if the DRR of each community is the only source of power for the community customers, the adequacy ratio will substantially decrease over time. In fact, even if the DRRs are sufficient to satisfy the customers' demand in long-term($\widehat{\mathcal{S}}(q,t)\ge s_q$), the system adequacy can not be guaranteed because of the white noise errors existed in the short-term forecasting scheme of demand and generation. Subsequently, we have to get help from the bulk generators located outside the community to cancel out the temporary noises and improve the adequacy ratio by generating extra energy \textit{on demand}.
\subsection{Canceling out the Noise by Energy Requests}
As mentioned before, if the value of $ \mathcal{S}(q,t) $ falls below some threshold ($ s_q-\lambda $), the system adequacy will be endanger. In this subsection, we use the bulk generation as a back-up plan to prevent such event. To do this, we design a controller to watch the amount of stored energy in different communities during the time. In the case that $ \mathcal{S}(q,t)\le s_q-\lambda $ for the $ q^{th} $ community, the controller asks the bulk generation to fill the gap and cancel out the noise $ -\mathcal{W}_t $ by providing the community with $ \mathcal{W}_t $ units of energy.

Here, we focus on what has to be done by the LLMU specified in Figure \ref{fig:storage}. The recently explained scenario which specified the functionality of LLMU cannot be implemented in the real-world as the values of $\mathcal{G}(q,t)$ and $\mathcal{D}(q,t)$ are obtained from random processes. This urges us to forecast theses values in short term (for example, every 15 minutes). Algorithm \ref{alg:decmakinprocess} shows the practical way of implementing LLMU using forecasting models for the $ q^th $ community. As mentioned before, by forecasting (estimating) the power demand and generation using ARIMA model, we will obtain white noise errors added to the real values of $\mathcal{G}(q,t)$ and $\mathcal{D}(q,t)$ as the estimated values. These noises on the estimated power values will add some errors in the form of Brownian Motion processes to the estimated values of stored energy.
\section{SUMMARY AND OUTLOOK}
This paper proposed a novel hybrid scheme for forecasting the power demand and generation in a residential power distribution network. Our forecasting scheme had two tiers: long-term demand/generation forecaster which is based on Maximum-Likelihood Estimator (MLE) and real-time demand/generation forecaster which is based on Auto-Regressive Integrated Moving-Average (ARIMA) model. The paper also showed how bulk generation improves the adequacy of our residential system by canceling-out the forecasters estimation errors which are in the form of Gaussian White noises.

{{\bf {Kianoosh G. Boroojeni}} is a PhD student of computer science at Florida International University. Kianoosh received his B.Sc degree from University of Tehran in 2012.  His research interests include reliability analysis, network design, and smart grids. He is co-author of a couple of books published by Springer and MIT Press.

~~{{\bf {Shekoufeh Mokhtari}} is a PhD student at Florida International University. She received her B.Sc. from University of Isfahan. Her research interests are in the general area of computer networking, including network security  and network measurement. \\

~~{{\bf {Mohammadhadi Amini}} was born in Shahreza (Qomsheh). He received the B.S. and M.Sc. degree in electrical engineering from Sharif University of Technology and Tarbiat Modares University in 2011 and 2013.respectively. He is currently working toward his Ph.D in Electrical Engineering at Electric Energy systems Group (EESG), Carnegie Mellon University. He is with SYSU-CMU Joint Institue of Engineering and SYSU-CMU Shunde International Joint Research Institute. His research interests include Plug-in Hybrid Electric Vehicles (PHEVs) optimization and control, smart distribution networks optimization, and state estimation in modern power system. \\

~~{{\bf {S. S. Iyengar}} is currently the Ryder Professor
and Director of the School of Computing \&
Information Sciences at the Florida International
University, Miami, FL since August 2011 and
before he was the Roy Paul Daniels Professor
and Chairman of the Computer Science Department
at Louisiana State University. During his
tenure at LSU he lead the Wireless Sensor
Networks Laboratory and the Robotics Research
Laboratory. He has also authored/coauthored
eight textbooks and edited 12 books in the areas
of Distributed Sensor Networks, Parallel Programming and Graph Theory,
published in CRC Press/Taylor and Francis/John Wiley/Springer-Verlag/
Prentice Hall/Chinese). He is a Fellow of the IEEE, Fellow of Association of
Computing Machinery (ACM), American Association for the Advancement
of Science (AAAS), Member European Academy of Science (EURASC),
and Fellow of The Society for Design and Process Science (SDPS). He is
a Golden Core member of the IEEE-CS and a recipient of the Lifetime
Achievement Award for Outstanding Contribution to Engineering Awarded
by Indian Institute of Technology, Banaras Hindu University, ICAM, 2012.
He has published over 400 articles, 40 keynote speeches, yearly
workshops at Raytheon, Army Research, and Indo-US Workshop for
sensor network, three patents and five patent disclosures pending,
supervised 50 PhD dissertations and led the new faculty hiring at FIU and
LSU. Further, Iyengar is the founding Editor-In-Chief of the International
Journal of Distributed Sensor Networks and has been an Associate Editor
for IEEE Transaction on Computers, IEEE Transactions on Data and
Knowledge Engineering, and guest Editor of IEEE Computer Magazine.
He has been an editorial member of many IEEE journals in advisory roles.
His research interests include Computational Sensor Networks (Theory
and Application) Parallel and Distributed Algorithms and Data Structures
Software for Detection of Critical Events Autonomous Systems
Distributed Systems Computational Medicine. G-Index: 50 and his
papers are cited more than 6000 times. \\\\
\end{document}